\documentclass{article}
\usepackage{ijcai24}

\usepackage[table]{xcolor}
\usepackage{times}
\usepackage{soul}
\usepackage{url}
\usepackage[hidelinks]{hyperref}
\usepackage[utf8]{inputenc}
\usepackage[small]{caption}
\usepackage{graphicx}
\usepackage{amsmath}
\usepackage{amsthm}
\usepackage{booktabs}
 \usepackage{multirow} 
\usepackage{algorithm}
\usepackage{algorithmic}
\usepackage[switch]{lineno}
\usepackage{soul, color, xcolor}
\usepackage{colortbl}
\usepackage{siunitx} % Align numbers at decimal point
\usepackage{array}   % Required for advanced table formatting
\usepackage{amssymb}  % For \checkmark
\usepackage{subfigure} % for \subfigure
\usepackage{footnote}

\usepackage{fancyhdr}

\begin{document}

\title{XNN: Paradigm Shift in Mitigating Identity Leakage within Cloud-Enabled Deep Learning}

% Single author syntax
\author{
Kaixin Liu$^{1}$\footnotemark[1]\footnotemark[2]
\and
Huixin Xiong$^{1}$\footnotemark[1]\and
Bingyu Duan$^{2}$\and
Zexuan Cheng$^{3}$\and
Xinyu Zhou$^{1}$\and \\
Wanqian Zhang$^{2}$\and
Xiangyu Zhang$^{1}$
\affiliations
$^1$MEGVII Technology\\
$^2$Institute of Information Engineering, Chinese Academy of Sciences, China\\
$^3$Beihang University, China\\
% \emails
% \{first, second\}@example.com,
% third@other.example.com,
% fourth@example.com
}

\maketitle

\renewcommand{\thefootnote}{\fnsymbol{footnote}}
\footnotetext[1]{Equal Contribution.}
\footnotetext[2]{Corresponding author. (e-mail: \href{mailto:liukaixin@megvii.com}{liukaixin@megvii.com})}

% 恢复脚注编号为数字
\renewcommand{\thefootnote}{\arabic{footnote}}

\begin{abstract}
    In the domain of cloud-based deep learning, the imperative for external computational resources coexists with acute privacy concerns, particularly identity leakage. 
To address this challenge, we introduce XNN and XNN-d, pioneering methodologies that infuse neural network features with randomized perturbations, striking a harmonious balance between utility and privacy.
XNN, designed for the training phase, ingeniously blends random permutation with matrix multiplication techniques to obfuscate feature maps, effectively shielding private data from potential breaches without compromising training integrity. 
Concurrently, XNN-d, devised for the inference phase, employs adversarial training to integrate generative adversarial noise. 
This technique effectively counters black-box access attacks aimed at identity extraction, while a distilled face recognition network adeptly processes the perturbed features, ensuring accurate identification.
Our evaluation demonstrates XNN's effectiveness, significantly outperforming existing methods in reducing identity leakage while maintaining a high model accuracy. 
\end{abstract}

\section{Introduction}

In recognition tasks, such as face and fingerprint recognition, dataset owners often encounter varying data distributions. 
This is particularly evident in face recognition, where datasets differ in aspects like illumination, angle, resolution, and identity. 
Fine-tuning a pre-trained model with the dataset owner's specific data can significantly enhance model accuracy. 
Commonly, dataset owners outsource their data to cloud platforms for this fine-tuning. 
However, if the cloud platform is not trustworthy, it risks complete privacy leakage of the dataset. 
Our paper addresses this issue of identity leakage in recognition datasets, using face recognition as a case study.

\begin{figure}
  \centering
  \includegraphics[width=.35\textwidth]{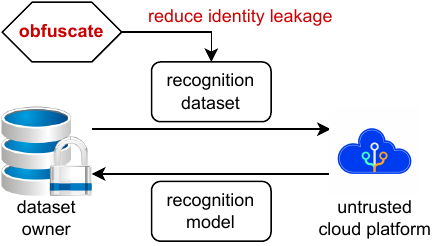}
  \caption{The framework of the training stage.}
  \label{fig:OutsourcedTraining}
  \vspace{-3mm}
\end{figure}

In the cloud-based training phase, outsourcing a dataset to an untrusted cloud platform poses a risk, as depicted in Figure~\ref{fig:OutsourcedTraining}. 
Simply making face images unrecognizable to humans does not suffice for privacy protection. 
If a dataset owner extracts features using a public pre-trained model and outsources these features, adversaries can still identify individuals by comparing these features with those from a large-scale gallery dataset. 
Thus, our focus on preventing identity leakage, which we define as the adversary's ability to match persons in the outsourced dataset.

To mitigate this risk, we propose a novel method, XNN (where 'X' signifies unknown and 'NN' stands for neural network). This involves applying random patch permutation and random matrix multiplication to the features of a pre-trained model. 
The obfuscation parameters are randomly initialized and kept secret from everyone except the dataset owner. We demonstrate that this obfuscation approach effectively reduces identity leakage to almost zero while maintaining high model accuracy.

\begin{figure}
  \centering
  \includegraphics[width=.45\textwidth]{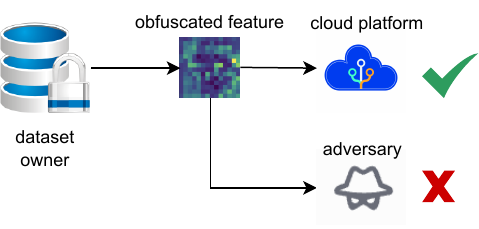}
  \caption{The framework of inference stage.}
  \label{fig:xnn-d}
  % \vspace{-3mm}
\end{figure}

In the inference stage of face recognition, the privacy concerns mirror those in the cloud-based training phase, as depicted in Figure~\ref{fig:xnn-d}. When data owners transmit locally extracted facial features to a server for recognition, these transmissions are vulnerable to interception by third-party attackers. Such attackers could exploit this data to infer identities. To counteract this threat, we introduce XNN-d, with 'd' denoting 'distillation', encapsulating the essence of our strategy.

Our methodology centers around a dual-stage training process, designed to craft a model that upholds privacy (comprising both the feature extractor for the data owner and the recognition network for the cloud server). The first stage involves adversarial training to create a generator of adversarial noise. This noise, when merged with features from a publicly pre-trained model, effectively thwarts the identity recognition efforts of unauthorized parties. However, this alone would impede the cloud server's recognition network from fulfilling its recognition role.
To circumvent this limitation, we incorporate a distillation technique. We train recognition network to recognize features that are modified by the noise. This innovative approach ensures that while the recognition attempts by potential attackers are rendered ineffective, the distilled recognition network continues to efficiently and accurately perform facial recognition tasks.

The XNN and XNN-d framework strikes a delicate balance, offering robust privacy safeguards without compromising the efficacy and accuracy of facial recognition, thereby aligning with both the security needs and functional demands of real-world applications.

% \begin{figure}
%   \centering
%   \includegraphics[width=.45\textwidth]{img/identity leakage.pdf}
%   \caption{The notion of identity leakage. The adversary has a collection of gallery images and the outsourced feature of a private face image, but does not have access to the private face image. The identity leakage is defined as the accuracy of finding the correct gallery image by the adversary.}
%   \label{fig:IdentityLeakage}
%   \vspace{-3mm}
% \end{figure}

The main contributions of this paper are summarized as follows:

(a)  We proposed an obfuscation method called XNN which can reduce the identity leakage to almost 0 with a relatively small drop in model accuracy. It outperforms other obfuscation-based methods in our specific problem setting.

(b) We developed XNN-d for the inference phase, integrating adversarial training and distillation to secure feature transmissions and maintain the accuracy of recognition network, effectively balancing privacy and functionality.

(c) We conducted extensive experiments to validate the effectiveness and robustness of XNN and XNN-d on various recognition datasets and transformer structures.

The rest of the paper is structured into several sections, starting with Section~\ref{sec:RelatedWorks}, which provides a review of the related literature in the privacy-preserving domain. Section~\ref{sec:Pipeline} and~\ref{sec:Pipeline_d} present the detailed description of the pipeline of our proposed models, including the obfuscation mechanism. The evaluation method is discussed in Section~\ref{sec:EvaluationMethod}, which outlines the identity leakage and defines the expectation attack. In Section~\ref{sec:ExperimetResults}, we present experimental results under the evaluation metric to validate the effectiveness of XNN. We then discuss the advantages and limitations of our proposed approach in Section~\ref{sec:Discussion}. Finally, a conclusion of our work is given in Section~\ref{sec:Conclusion}, summarizing our contributions, highlighting the limitations of our work, and suggesting future research directions.

\section{Related works}
\label{sec:RelatedWorks}

In the field of instance transformation-based obfuscation, various methods have been explored, each with its unique approach and level of effectiveness. XNN and XNN-d, our proposed methods, are significant contribution to this domain, building upon and advancing previous works.

InstaHide~\cite{huang2020instahide} obfuscates dataset with random sign flip.
It first picks up several images from the private dataset, and picks up the same number of images from a public dataset.
Then the private and public images are linearly combined with random positive coefficients that sum up to 1.
At last, a pixel-wise random sign flip is applied to the combined image.
This random sign flip is used as one-time-key, and will be different for each instance.
However, this seemingly reasonable obfuscation has already been broken by reconstruction attacks~\cite{carlini2020private}.

Dauntless \cite{xiao2021dauntless} replaced the sign flip operation in InstaHide with a random neural network.
It applies the obfuscation neural network on the original image directly.
The parameters of the obfuscation network are only known to the dataset owner.
This obfuscation provides stronger privacy protection than sign flip.
Dauntless conducted experiments on small tasks.

TaskAugment \cite{xiao2021art} concatenates multiple instances from different tasks into a single instance.
For example, given a $c$-classification problem with input $x$ and output $y$, and another $c'$-classification problem with input $x'$ and output $y'$, TaskAugment defines a new $c \times c'$-classification problem with input $(x, x')$ and output $(y, y')$.
The obfuscation used in TaskAugment is the same with Dauntless, i.e. random neural network obfuscation on the original image. %
% TaskAugment also proposed several improvements to the design of the obfuscation function and training strategies on the obfuscated data.
New architectures of the obfuscation neural network have been proposed to replace the fully-connected network in Dauntless, which made the training on obfuscated data easier for convolution networks or vision transformers.
Transfer learning technique is also discussed in TaskAugment for accuracy improvement. %
Obfuscation applied to the features of a pre-trained model's first few layers achieves better model accuracy than the obfuscation on the original image. %

NeuralCrypt~\cite{yala2021neuracrypt} focus on the private learning of health data MIMIC-CXR~\cite{johnson2019mimic} and CheXpert~\cite{irvin2019chexpert} (X-ray images), which are relatively large datasets with real privacy concerns.
It encodes raw patient data using a random neural network, and publishes both the encoded data and associated labels publicly.
The obfuscation network is a composition of random convolution and random patch shuffle.
The author of NeuralCrypt admits that their theoretical results do not guarantee the security of their specific obfuscation network.
Therefore, they provided empirical experiments to exhibit its resistance under several attacks, as well as a challenge dataset for public attack tests.
So far, no attack has broken NeuralCrypt under the circumstance that no original data is available to the adversary.
The author of NeuralCrypt also released an easier challenge, where the original data is available to the adversary (although real adversaries are not supposed to have this ability).
In this easier scenario, a complete break has been achieved by~\cite{carlini2021neuracrypt}.

% Syfer? Dauntless?

XNN draws inspiration from the "Image and Model Transformation" concept~\cite{kiya2022image}, where vision transformer (ViT) models are trained on encrypted images created via random patch permutation and other techniques. We discovered that ViT models could be directly trained on encrypted images, maintaining accuracy without knowledge of the secret key.
Separately, we propose the XNN-d framework, which is inspired by~\cite{wu2023backdoor}. This approach first introduces a backdoor attack to effectively mitigate bias, and then utilizes model distillation to eliminate the security threats posed by the backdoor attack itself while preserving the debiasing effect. Specifically, XNN-d mitigates privacy leakage through adversarial noise and subsequently improves the recognition accuracy of the student model via distillation, thus ensuring privacy protection without compromising the model's ability to perform recognition tasks.

\section{Pipeline of XNN}
\label{sec:Pipeline}
\begin{figure}
  \centering
  \includegraphics[width=.45\textwidth]{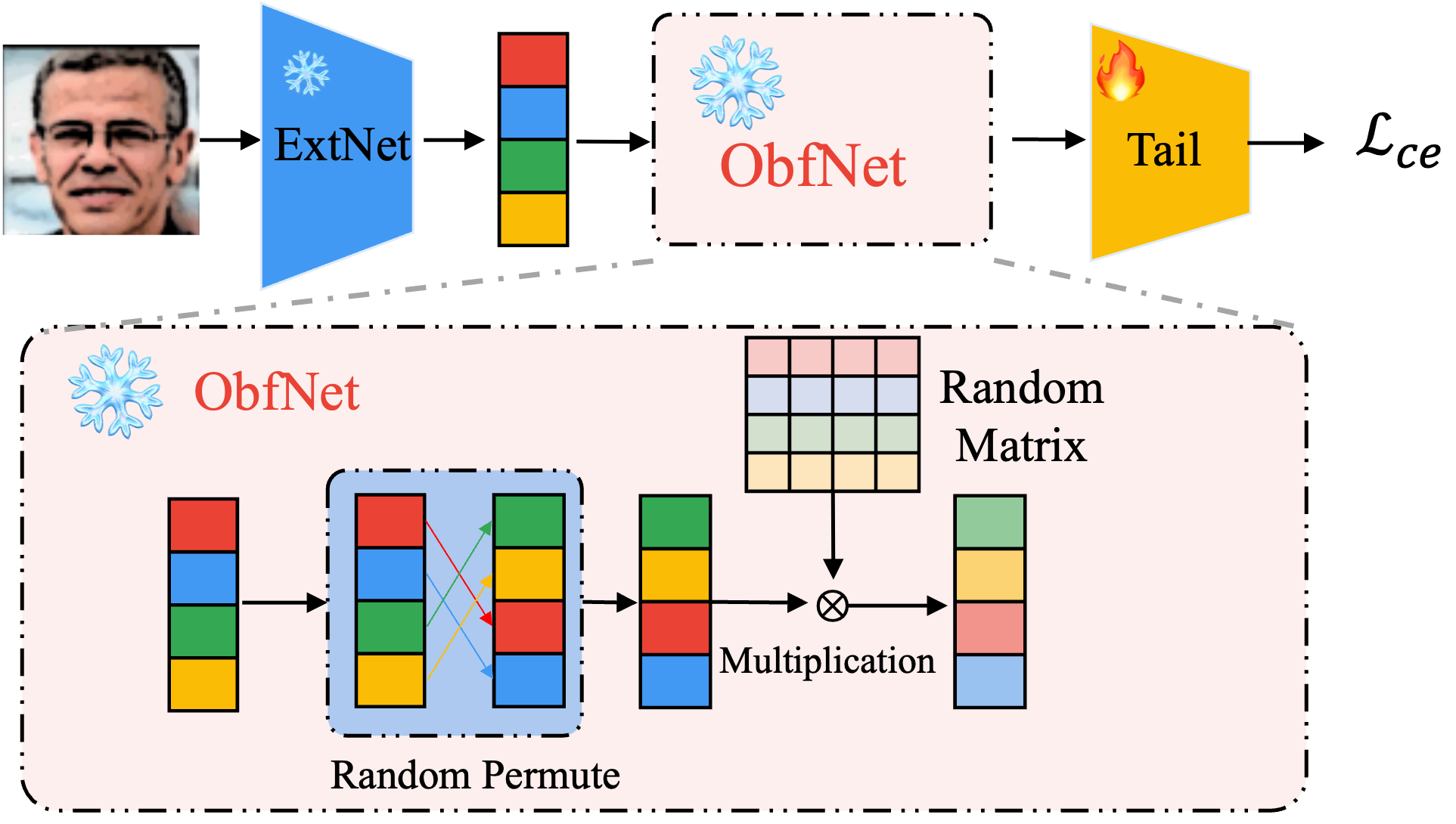}
  \caption{The pipeline of XNN.}
  \label{fig:XNNppl}
  \vspace{-3mm}
\end{figure}

Our XNN framework, depicted in Figure~\ref{fig:XNNppl}, integrates a pre-trained feature extractor (ExtNet), an obfuscation network (ObfNet), and a recognition network (RecNet). Initially, the dataset owner's images pass through ExtNet and ObfNet for transformation. The cloud platform then trains RecNet on these obfuscated features and anonymous labels, before returning the trained RecNet to the owner for final model assembly and inference.

\subsection{Instance obfuscation}
In the instance obfuscation stage, ExtNet, trained on a public dataset, processes each image into a feature map. The output of ExtNet alone, while visually unrecognizable, remains susceptible to identity leakage, especially if ExtNet's details are known to potential adversaries. To counteract this, ObfNet applies a combination of random patch permutation and matrix multiplication to ExtNet's output. The obfuscation parameters, known only to the dataset owner, prevent adversaries from reconstructing the original image.

\subsection{Training on the obfuscated data}
After obfuscation, the dataset is sent to the cloud platform for model training.
The cloud platform trains the RecNet with the obfuscated feature as input, and its anonymous label as supervising signal.
The anonymous label is generated by mapping the original label (e.g. names) to a random label (e.g. numbers).
The mapping is one-to-one and is only known to the dataset owner.

\subsection{Inference on the obfuscated query}

Once the cloud platform finishes training the RecNet on the obfuscated data, the RecNet is sent back to the dataset owner.
To get the recognition feature vector of a new face image, the dataset owner needs to successively apply the ExtNet, ObfNet, and RecNet to the image.
Because the RecNet is trained on the obfuscated data, and the parameters of the ObfNet are only known to the dataset owner, this RecNet can only be used by this dataset owner.
Other dataset owners can not obtain accuracy increasement with the help of the RecNet, even if their data distributions are similar.
They do not know the parameters of the ObfNet, therefore they can not transform their face images to the corresponding obfuscated feature for inference.
\begin{figure}
  \centering
  \includegraphics[width=0.5\textwidth]{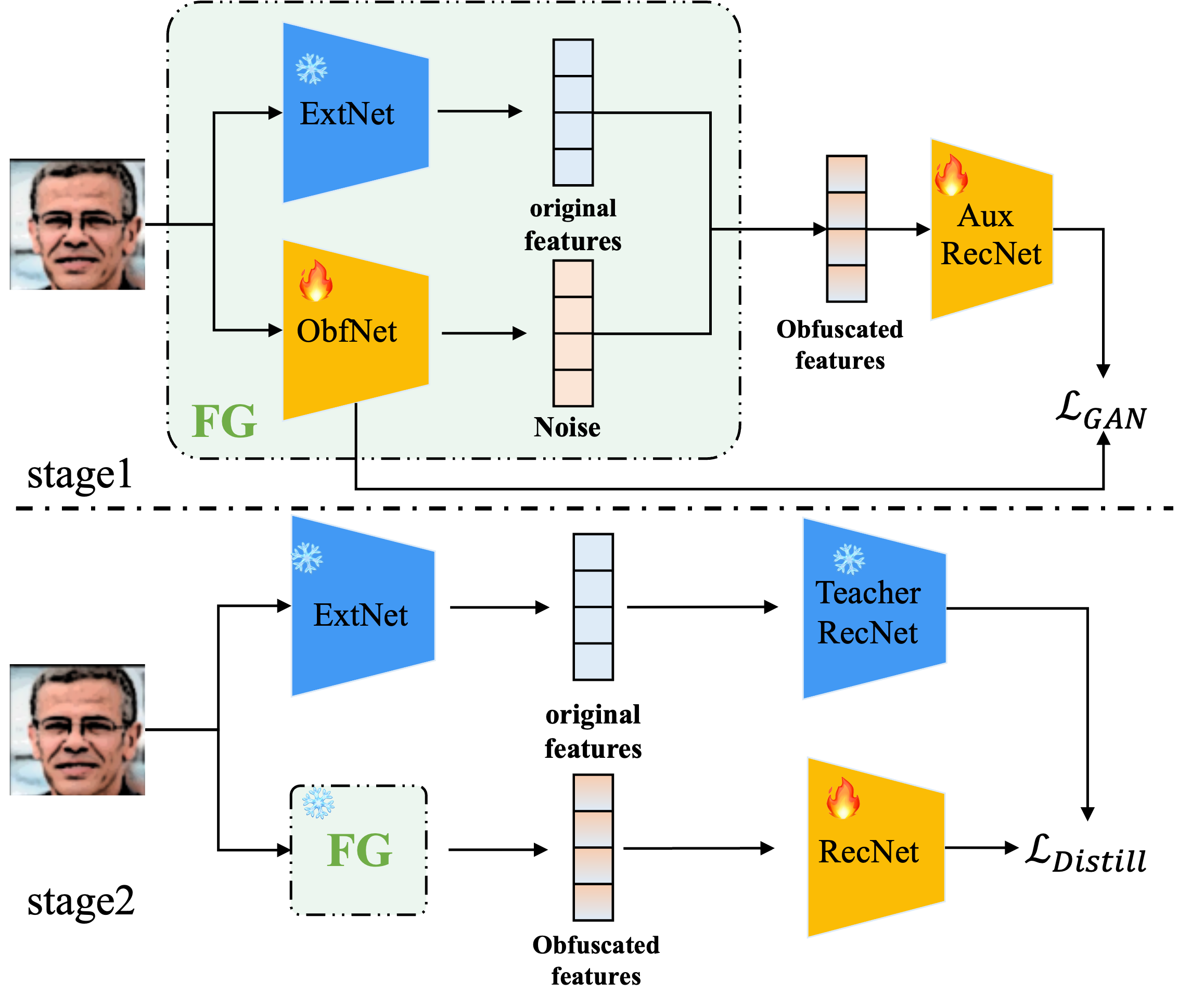}
  \caption{The pipeline of XNN-d}
  \label{fig:XNN-d}
  \vspace{-2mm}
\end{figure}
\section{Pipeline of XNN-d}
\label{sec:Pipeline_d}
XNN-d is depicted in Figure~\ref{fig:XNN-d}, integrates a pre-trained feature extractor (ExtNet), an obfuscation network (ObfNet) and a recognition network (RecNet). Feature Generator (FG) is consist of ExtNet and ObfNet.

\subsection{The training of ObfNet}
In the first training stage, as show in Figure~\ref{fig:XNN-d}, we employ adversarial training techniques to train the noise encoder.
Specifically, we employ ExtNet and ObfNet to extract the original features and noise and obfuscated features is generated by mixing original features and noise.
RecNet is utilized to recognize the obfuscated feature.
By utilizing the Adversarial loss $\mathcal{L}_{GAN}$,the noise encoder is optimized to generate noised feature against the RecNet.

\subsection{Distillation of RecNet}
As show in Figure~\ref{fig:XNN-d}, to ensure the availability of the face recognition network, we employ Teacher RecNet to guide the learning of RecNet of server-side.
Specifically, we utilize ExtNet to extract clean original feature which is input to Teacher RecNet to acquire clean logits.
Also, obfuscated features is obtaining by FG and logits is acquiring by RecNet of server-side.
The distillation loss $\mathcal{L}_{distill}$ is utilized to empower the RecNet of server-side with the ability to recognize the obfuscated features.

\section{Evaluation method}
\label{sec:EvaluationMethod}

\subsection{Identity leakage}

The target of XNN is reducing the identity leakage of the obfuscated dataset under adversarial attacks.
Given an obfuscated feature (denoted as $f$), the adversary tries to identify whom this feature belongs to (denoted as $p$).
Suppose the adversary has a large face dataset (denoted as $B$), for example, a large public dataset.
If $B$ is large enough, another face image of $p$ (denoted as $b_p$) may happen to be contained in $B$.
Suppose for any person $p$ in the owner's dataset (denoted as $D$), there is another image of $p$ in $B$ (this is a quite strong enhancement of the adversary).
Then the identity leakage can be quantified as the accuracy of finding $b_p$ in $B$ given $f$, evaluated on all $p$ in $D$.
If we denote instance obfuscation from $p$ to $f$ as $f=\text{obf}(p)$, and the attack of finding an image $a$ in $B$ given $f$ as $a=A(f,B)$, then the identity leakage is quantified as
\begin{align}
Leak={\mathbb E}_{p\in D}[A(\text{obf}(p),B)==b_p] .
\end{align}

If there is no identity leakage, then $Leak$ will be $1/N$ (random guess), where $N$ is the size of dataset $B$.
And a total leakage of identity will lead to $Leak \to 1$.

\begin{figure}
  \begin{center}
  \includegraphics[width=.43\textwidth]{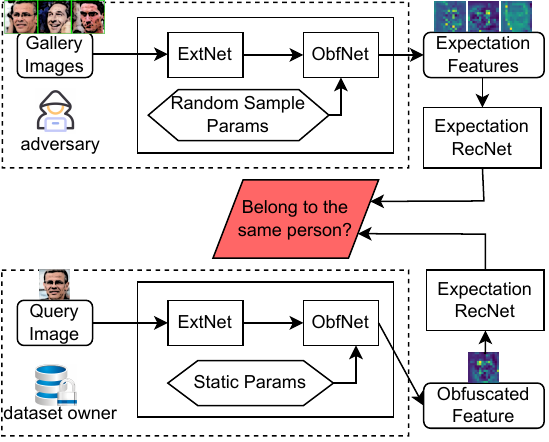}
  \end{center}
  \vspace{-2mm}
  \caption{The diagram of the expectation recognition attack.}
  \label{fig:AttackPPL}
  \vspace{-3mm}
\end{figure}

\subsection{Expectation attack}
To measure XNN's identity leakage, we devised an adversarial attack termed 'expectation recognition', depicted in Figure~\ref{fig:AttackPPL}. This attack aims to deduce the recognition feature vector from obfuscated features without knowing the dataset owner's obfuscation parameters. The adversary trains an Expectation-RecNet using expectation features, which employ the same ExtNet and a variant of ObfNet, with each batch receiving new obfuscation parameters. This forces the Expectation-RecNet to learn to identify recognition feature vectors regardless of obfuscation.

During the attack phase, Expectation-RecNet processes the obfuscated feature, either from the dataset owner or generated via expectation obfuscation, and outputs a recognition feature vector. The adversary then matches this vector against those from a large gallery dataset to find the most similar one, reflecting the top-1 accuracy.

In practical scenarios, the adversary's gallery dataset should be extensive, ideally encompassing images from the owner's dataset. 
For simplicity, our experiments use the test set of the owner's dataset as the gallery, significantly aiding the adversary. 
This setup allows for a straightforward comparison of recognition feature vectors, quantifying identity leakage through top-1 accuracy.

\subsection{Black-Box attack}
In our assessment of XNN-d, we adopted a nuanced approach known as black-box access attacks. In this context, the attacker is precluded from direct insight into ExtNet's inner workings, such as discerning the model's weights or its architectural nuances. However, they can subtly infer the model's behavior by strategically inputting data and scrutinizing the resultant outputs. This indirect probing facilitates the creation of a supplementary dataset, replete with features and labels, derived from a carefully curated gallery dataset. Upon this constructed dataset, the attacker then meticulously trains an analogous version of RecNet. While this iteration of RecNet is refined based on the auxiliary dataset, its efficacy hinges on the diversity of the gallery dataset. A well-varied gallery dataset empowers this trained RecNet to generalize with reasonable accuracy to the dataset utilized by the data owner in the inference phase, encapsulating the essence of ID information extraction in black-box access attacks.

\section{Experiment results}
\label{sec:ExperimetResults}
\subsection{Experimental Setup}

\begin{table}[b]
    \centering
    \resizebox{\columnwidth}{!}{
        \begin{tabular}{lrrrr}
            \toprule
            Dataset & Images & \#IDs & Trainset & Testset \\ 
            \midrule
            CelebA~\cite{liu2015faceattributes} & 203K & 10K & 177K & 20K \\ 
            MSRA~\cite{TrillionPairs} & 3,923K & 87K & 3,869K & 20K \\ 
            CASIA-WebFace~\cite{yi2014learning} & 456K & 11K & 394K & 20K \\ 
            VGGFace2~\cite{Cao18} & 3,142K & 9K & 2,778K & 20K \\ 
            \midrule
            IMDb-Face~\cite{WangCLHCQL18} & 961K & 40K & 902K & 20K \\ 
            FaceScrub~\cite{NgW14} & 32K & 0.5K & 25K & 2K \\ 
            \bottomrule
        \end{tabular}
    }
    \caption{Statistics of dataset information for evaluation.}
    \label{tab:datasets_info}
\end{table}

\noindent\textbf{Dataset.}We adopt six face datasets in our evaluation, including CelebA, CASIA-WebFace, FaceScrub, MSRA, VGGFace2, and IMDb-Face. 
For instance, the CelebA dataset is a large-scale face attributes dataset with more than 200K celebrity images collected from the internet and contains 10K identities.
And MSRA contains 3,923K aligned images cleaned from the MS-Celeb-1M dataset, while celebrity is the Asian dataset with 87K IDs.
The detailed information on the datasets involved in our evaluation experiments is shown in Table~\ref{tab:datasets_info}.
In particular, we divides the datasets according to Identities(IDs). 
We selected 1000 persons (100 persons for FaceScrub) from the original dataset, and each person samples 20 images as the test set. The test set does not intersect with the people in the training set.
The number of images of some IDs selected for testing is greater than 20, then the extra part will not be used in training and testing, so the total images used for the training and test sets will be slightly less than the original dataset.

\begin{table}
\centering
\begin{tabular}{ 
    >{\centering}p{0.2\linewidth} 
    S[table-format=2.2] 
    S[table-format=2.2] 
    S[table-format=2.2] }
    \toprule
    \multirow{2}{*}{Methods} & \multicolumn{1}{c}{\multirow{2}{*}{Utils ($\uparrow$)}} & \multicolumn{2}{c}{ASR ($\downarrow$)} \\
    & & {Facescrub} & {IMDB} \\
    \midrule
    Vanilla+PE & \textbf{67.77} & 67.77 & 67.77 \\
    IH(k=2) & 50.74 & 8.07 & 57.87 \\
    IH(k=3) & 25.33 & 6.37 & 33.52 \\
    NC & 39.67 & 9.39 & 32.48 \\
    \rowcolor{gray!25} XNN & 64.96 & \textbf{0.17} & \textbf{0.19} \\
    \bottomrule
\end{tabular}
\caption{Performance Comparison (\%) on CelebA.}
\label{tab:comparison_celeba}
\end{table}
\begin{table}
\centering
\begin{tabular}{ 
    >{\centering}p{0.2\linewidth} 
    S[table-format=2.2] 
    S[table-format=2.2] 
    S[table-format=2.2] }
    \toprule
    \multirow{2}{*}{Methods} & \multicolumn{1}{c}{\multirow{2}{*}{Utils ($\uparrow$)}} & \multicolumn{2}{c}{ASR ($\downarrow$)} \\
    & & {Facescrub} & {IMDB} \\
    \midrule
    Vanilla+PE & \textbf{53.33} & 53.33 & 53.33 \\
    IH(k=2) & 8.18 & 2.06 & 2.72 \\
    IH(k=3) & 3.03 & 1.68 & 1.53 \\
    NC & 37.05 & 2.38 & 6.17 \\
    \rowcolor{gray!25} XNN & 50.03 & \textbf{0.18} & \textbf{0.16} \\
    \bottomrule
\end{tabular}
\caption{Performance Comparison(\%) on WebFace.}
\label{tab:comparison_webface}
\end{table}

In the setting of XNN and XNN-d, FaceScrub and IMDB are regarded as the auxiliary dataset of adversary for attacking, while CelebA and CASIA-WebFace are used in XNN's and XNN-d privacy training process.
The rest of the datasets are used to evaluate the effectiveness of XNN and XNN-d on different datasets, which will be described in section~\ref{subsec:eval_datasets}.
Except for Facescrub's test set, which is 20 images for 100 people each, we randomly divide the original dataset into a training set and testing set consisting of 20,000 images of 1,000 people (20 per person).
For convenience, all the images are resized to 112$\times$112.

% \begin{table}
% 	\centering
%     \renewcommand\arraystretch{1.0}
%     \resizebox{\columnwidth}{!}{
%     % \setlength{\tabcolsep}{15pt}
% 	\begin{tabular}{l|rrrr}
% 		\toprule
%            Dataset & Images & \#IDs & Trainset & Testset \\ 
%         \midrule
%         CelebA~\cite{liu2015faceattributes} & 203K    &   10K  &  177K  & 20K    \\ 
%         MSRA~\cite{TrillionPairs}     & 3,923K    & 87K   & 3,869K  & 20K    \\ 
%         CASIA-WebFace~\cite{yi2014learning}    &  456K   &  11K  & 394K  &  20K    \\ 
%         VGGFace2~\cite{Cao18}       &  3,142K   &  9K  &  2,778K  &  20K   \\ 
%         \midrule
%         IMDb-Face~\cite{WangCLHCQL18}      & 961K  &  40K & 902K   &  20K  \\ 
%         FaceScrub~\cite{NgW14}          &  32K    & 0.5K   & 25K   &  2K   \\ 
% 		\bottomrule
%         % \hline
% 	\end{tabular}
%     }
%     \vspace{1mm}
%     \caption{Statistics of dataset information for evaluation. Top 4 commonly used face image datasets serve as privacy data. In addition, we included two auxiliary datasets, one large and one small, to serve as a reference for the adversary.}
%     \label{tab:datasets_info}
%     \vspace{-3mm}
% \end{table}

\noindent\textbf{Training details.}
To assess the effectiveness of our framework, we benchmark it against a non-private original model and several privacy-preserving techniques. 
Among these methods, NeuraCrypt~\cite{yala2021neuracrypt} provides a comprehensive analysis, and InstaHide~\cite{huang2020instahide} can be seamlessly integrated into existing distributed learning pipelines.

Under our setting, We take 5-blocks ViT-B~\cite{DosovitskiyB0WZ21} with MLPHead removed as the feature extractor, which is pre-trained by the self-supervised learning technology MoCo-v3~\cite{chen2021empirical}, and 6-blocks ViT-B with position embedding removed as the recognition network. 
For the NeuraCrypt, we follow the setting in~\cite{yala2021neuracrypt}, i.e., implementing data obfuscation with CNN-based encoder, which contains a position embedding module and patch shuffling module, and training the Vision Transformer that is invariant to patch orders. For simplicity, we set the same ViT-B as XNN to complete the recognition task.
The InstaHide privacy-preserving approach does not achieve data obfuscation through obfuscation networks like XNN or NeuraCrypt, but instead directly manipulates face image by sign flips on pixel values and mix up data augmentation method, which is to take its linear combination with $ k-1 $ randomly chosen images from the training set.

\begin{table}
	\centering
    \renewcommand\arraystretch{1.0}
    \resizebox{\columnwidth}{!}{
    \setlength{\tabcolsep}{15pt}
	\begin{tabular}{p{0.7cm}<{\centering}p{1.06cm}<{\centering}|c|cc}
% \hline
\toprule
\multirow{2}{*}{}        & \multirow{2}{*}{} & \multirow{2}{*}{Utility($\uparrow$)} & \multicolumn{2}{c}{ASR($\downarrow$)} \\
                         &                   &                          & Facescrub    & IMDB     \\ \midrule
\multirow{2}{*}{CelebA}  & baseline          & 70.49                    & 39.58        & 78.63    \\
&  \cellcolor{gray!25}XNN-d            & \cellcolor{gray!25} \textbf{73.15}                   &  \cellcolor{gray!25}\textbf{19.07 }       &  \cellcolor{gray!25}\textbf{21.02}   \\ \midrule
\multirow{2}{*}{WebFace} & baseline          & 52.89                    & 15.14        & 40.24    \\
                    & \cellcolor{gray!25} XNN-d             & \cellcolor{gray!25}\textbf{ 53.61 }                   &  \cellcolor{gray!25}\textbf{1.44}         & \cellcolor{gray!25} \textbf{13.04 }  \\ 
      \bottomrule % \hline
\end{tabular}
 }
 \vspace{1mm}
 \caption{Performance Comparison of XNN-d (\%) on CelebA and WebFace Datasets.}
 \label{tab:comparison_XNN-d}
 \vspace{-3mm}
\end{table}
For convenience and fair comparison, it is feasible to transfer XNN's base architecture without an obfuscation module to InstaiHide.
And XNN-d utilizes a 5-block ViT-B with LN as the feature extractor, and the noise encoder shares the same structure as the feature extractor. Additionally, a 6-block ViT-B with position embedding removed serves as the recognition network.
% Due to we are the first to explore the scenario which trusted servers provide face recognition service while preventing identity leakage. For the same obfuscated features, trusted servers can identify them, but other unauthorized third parties cannot.
We are the first to explore the scenario where trusted servers provide face recognition services while safeguarding against identity leakage. 
In this setting, trusted servers can recognize obfuscated features, yet these same features remain undecipherable to unauthorized third parties.
Therefore, in our experiments, we only compare it with the baseline which remove the noise encoder.
% The training of XNN-d consists of two stages. In the first stage, we employ  adversarial training techniques to train the noise encoder. In the second stage, we employ distillation techniques to train the recognition network.

During the training of the above model, the SGD~\cite{robbins1951stochastic} optimizer with Nesterov momentum~\cite{pmlr-v28-sutskever13} is used with the learning rate and weight decay term to be 0.05 and 0.00004, respectively.
In addition, we choose CosineAnnealingLR~\cite{loshchilov2016sgdr} as the learning rate scheduler.
The batch size for training all of these models is 256.
To avoid dominant leakage of any single image in InstaHide, we set the conservative upper threshold to 0.65 following~\cite{huang2020instahide}.
% GAN 的系数

\subsection{Comparison with Existing Privacy-Preserving Approaches}
\label{subsec:eval_comparison}

Table~\ref{tab:comparison_celeba}, Table~\ref{tab:comparison_webface} summarizes the performance comparison of XNN for the CelebA and WebFace dataset, while Table~\ref{tab:comparison_XNN-d} show XNN-d's.
As for XNN, the Vanilla model, which consists of ExtNet and RecNet without any modifications, is used as the backbone. The Pretrained Ext model (PE) denotes pre-training PE by the self-supervised learning technology MoCo-v3, while the InstaHide (IH) model and the NeuraCrypt (NC) model represents the Vanilla model with corresponding privacy protection methods, respectively.
In XNN-d, we added Layer Normalization (LN) after the XNN backbone's ExtNet as the baseline.
It should be noted that the Utility Accuracy (Util) and Attack Success Rate (ASR) are used as performance metrics. The former measures the accuracy of the model in identifying faces, while the latter measures the ability of an attacker to obtain sensitive information from the model.

\begin{table}
	\centering
    \renewcommand\arraystretch{1.0}
    \newcommand{\tabincell}[2]{\begin{tabular}{@{}#1@{}}#2\end{tabular}}
    \resizebox{\columnwidth}{!}{
	\begin{tabular}{cccc|cc}
		\toprule
		Vanilla & \tabincell{c}{Pretrain \\ Ext} & \tabincell{c}{With \\ Obf} & \tabincell{c}{Pretrain \\ Rec} & Utils($\uparrow$) & ASR($\downarrow$)\\
		\midrule
		\checkmark &    &  &  & 26.11 & 26.11 \\
  		\checkmark & \checkmark   &  &  & 67.77 & 67.77 \\
		\checkmark & \checkmark   & \checkmark &  & 64.96 & 0.17 \\
        \checkmark & \checkmark   &  & \checkmark & 7.95 & 7.95 \\
		\bottomrule
        % \hline
	\end{tabular}
    }
\vspace{1mm}
\caption{Ablation Studies(\%) on different modules of XNN on CelebA dataset.}
\label{tab:ablation_na}
% \vspace{-3mm}
\end{table}

In XNN, the original model(i.e., Vannila+PE) achieves the highest accrucy for face recognition, however the absence of perturbation mechanism leads to an severe risk of identity leakage.
Other privacy-preserving methods could provide privacy protection to some extent, but it's not such effective as ours and significantly decrease the model accrucy(e.g., from 67.77\% to 25.33\% in Table~\ref{tab:comparison_celeba}).
In contrast, our proposed method XNN provides the best privacy protection while only slightly decrease model accuracy.
Regarding utility, XNN preserved 95.85\% of the model accuracy on the CelebA dataset (from 67.77\% to 64.96\%), and maintained an average model accuracy of 96.89\% on four privacy datasets, which is shown in Table~\ref{tab:comparison_celeba}, Table~\ref{tab:comparison_webface} and Table~\ref{tab:diff_dataset}.
Notably, XNN achieves an ASR of 0.17\% on the Facescrub and IMDb-Face, which is significantly lower than all other methods and closely align with those of the random conjecture.
XNN was found to significantly reduce the average identity leakage by 99.77\% on four privacy datasets, including CelebA, WebFace, VGGFace2, and MSRA.

In XNN-d, comparing to baseline, our method effectively reduces the attack success rate (ASR) while maintaining good usability. For instance, on Celeba dataset, XNN-d decrease ASR by 20.51\% and 57.61\% for attacker dataset Facescrub and IMDB respectively. Surprisingly, under the influence of distillation techniques, it even improved Utility by 2.66\% and 0.72\% on CelebA and WebFace datasets.
% \begin{table}
% 	\centering
%     \renewcommand\arraystretch{1.0}
%     \resizebox{\columnwidth}{!}{
%     \setlength{\tabcolsep}{20pt}
% 	\begin{tabular}{c|cc}
% 		\toprule
% 		Methods & Utils($\uparrow$) & ASR($\downarrow$)\\
% 		\midrule
% 		only RP & 67.37 & 33.14 \\
% 		only RMM   & 66.58 & 0.19 \\
%   		RP \& RMM   & 64.96 & 0.17 \\
% 		\bottomrule
%         % \hline
% 	\end{tabular}
%     }
% \vspace{1mm}
% \caption{Ablation Studies(\%) on the Different Obfuscating Methods on the CelebA. This study evaluates the effectiveness of two obfuscation techniques, Random Permutation (RP) and Random Matrix Multiplication (RMM), included in ObfNet.}
% \label{tab:ablation_obf}
% % \vspace{-3mm}
% \end{table}

The comparison of performance reveals that XNN models exhibit comparable utility scores to the Vanilla model while demonstrating significantly lower ASR scores. 
On the other hand, while NC and IH models have shown outstanding results on attribute classification tasks, they fall behind XNN in terms of both utility and privacy on face recognition tasks. 
XNN-d ensures an enhancement in privacy protection without compromising utility in the inference stage of face recognition.
This highlights the pressing need for more advanced methods to tackle these challenges, especially considering the larger class space and strong perturbations involved in face recognition tasks, which make it challenging for neural networks to achieve accurate ID alignment.
This outcome suggests that XNN and XNN-d models achieve a favorable balance between privacy and utility, highlighting their potential as a promising solution for privacy-preserving face recognition. 

\subsection{Ablation Studies}
\label{subsec:eval_ablation}

% This study proposes Random Permutation (RP) and Random Matrix Multiplication (RMM) as two obfuscating mechanisms that balance the trade-off between privacy and accuracy in deep learning models. In order to evaluate the effectiveness of these proposed methods, we conducted ablation studies on different obfuscating techniques using the CelebA dataset, as shown in Table~\ref{tab:ablation_obf}.

% Our results demonstrate that models using only RP or only RMM achieve higher utility accuracy compared to models using our proposed method with both RP and RMM. However, such models also show an increased risk of ID leakage. These findings suggest that the combination of RP and RMM can effectively enhance privacy protection while maintaining high model accuracy.

\noindent\textbf{Analysis on different XNN's module.} 
In XNN model, we employ the obfuscation mechanism to balance the trade-off between privacy and accuracy in deep learning models.
In addition to incorporating an obfuscation mechanism, our approach pre-trains ExtNet to avoid involving the user-side model in the backpropagation during the fine-tuning stage, as compared to the Vanilla model. 
In this study, we analyze the effectiveness of the obfuscation mechanism and whether ExtNet and RecNet benefit from using pre-training weights.
% We evaluate the impact of pre-trained ExtNet on our proposed scheme. 
% Furthermore, we performed an ablation study to analyze why the RecNet  of the network architecture does not require pre-training. 
% Furthermore, we analyze why the RecNet  of the network architecture 
% Specifically, we compared the performance of RecNet with and without pre-training to demonstrate the effectiveness of our approach.

\begin{figure}
  \centering
	\subfigure{
    \includegraphics[height=3.8cm]{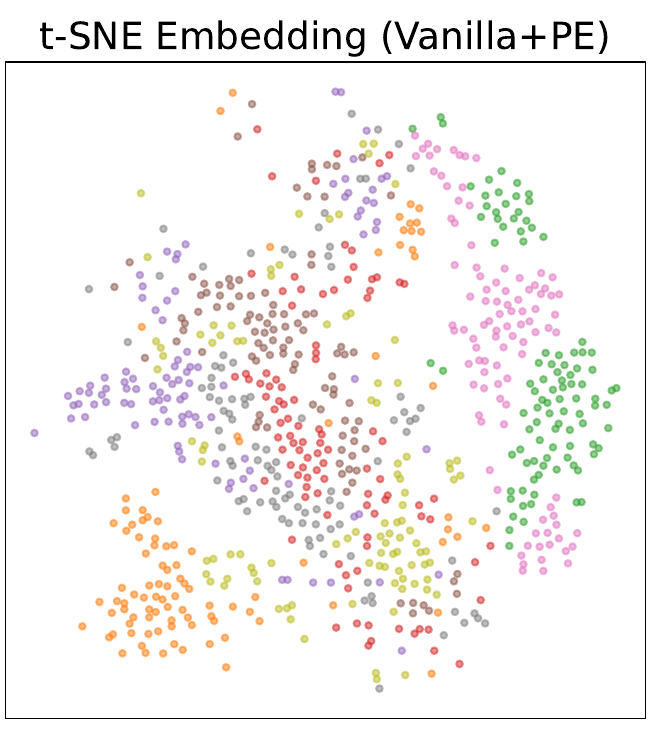}
		\label{fig:tsne_vanilla}
	}
    % \vspace{-3mm}
 	\hspace{-3mm}
	\subfigure{
	\includegraphics[height=3.8cm]{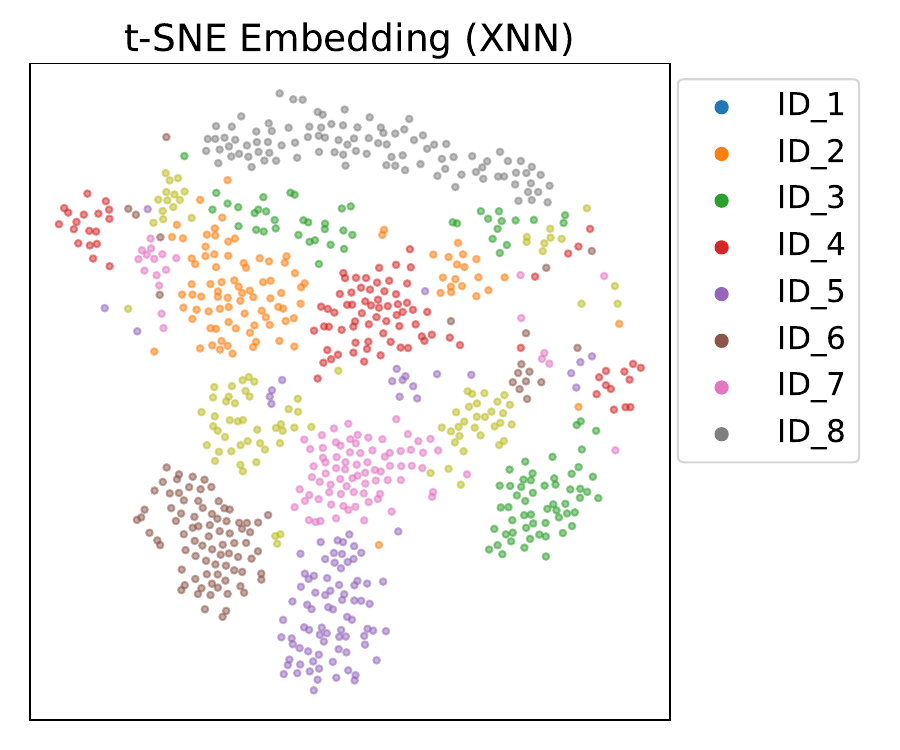}
			\label{fig:tsne_obf}
	}
  \caption{Visualization of the features before and after obfuscation on MSRA. Different colors represent samples from different persons.}
  \label{fig:tsne}
  \vspace{-3mm}
\end{figure}

\begin{figure}
  \centering
	\subfigure{
    \includegraphics[height=3.6cm]{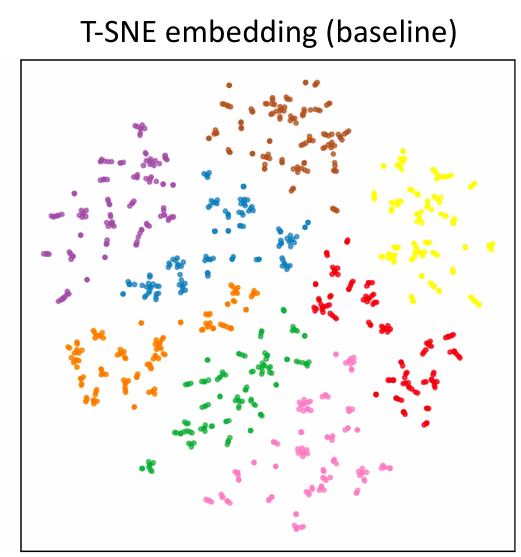}
		\label{fig:baseline-tsne}
	}
    % \vspace{-3mm}
 	\hspace{-3mm}
	\subfigure{
	\includegraphics[height=3.6cm]{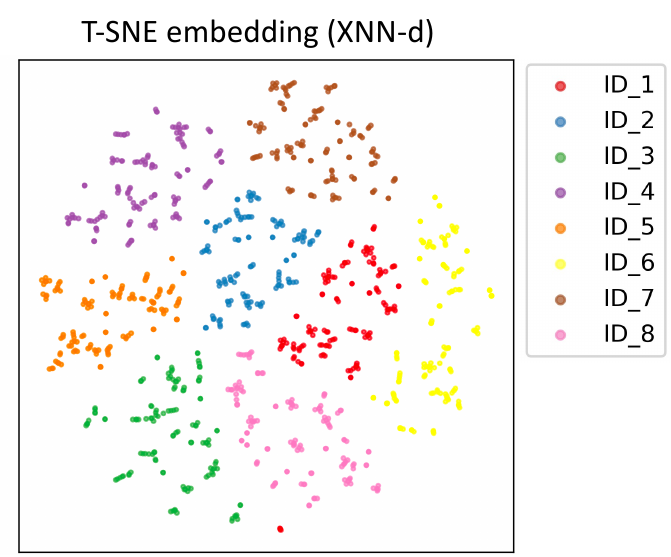}
			\label{fig:xnnd-tsne}
	}
  \caption{Visualization of the features of baseline and XNN-d on MSRA. Different colors represent samples from different persons.}
  \label{fig:tsne2}
  \vspace{-3mm}
\end{figure}

As shown in Table~\ref{tab:ablation_na}, although the obfuscation mechanism slightly reduces usability (from 67.77\% to 64.96\%), it significantly enhances the security of the model (from 67.77\% to 0.17\%).
Additionally, compared to the baseline, employing pre-trained ExtNet and freezing it improves utility by 41.66\%. 
This means that clients can avoid model training, thereby reducing the resource demands on the client side.
When RecNet uses pre-trained parameters and freeze it, the model's performance decreases meaning that server-side training is inevitable.
% This is because the deep layers of the model extract dataset-specific features, and employing pre-trained parameters hinders the model's training.
% In Table~\ref{tab:ablation_na}, we present the results of our ablation study on the network architecture for our approach on CelebA dataset. 
% We evaluate the performance of our approach under four different configurations.
% We observe that pre-training of ExtNet improves the performance, and adding obfuscation leads to acceptable drop in accuracy with significant enhanced privacy protection. 
% Additionally, the results suggest that pre-training of RecNet is not necessary in our approach. 
% We achieve the best utility of 67.77\% with pre-trained ExtNet and ASR of 0.17\% with obfuscation.

\noindent\textbf{Visualizing the Features of Faces.}
To figure out how XNN and XNN-d achieved such a balance, we visualized the original and obfuscated features by t-SNE~\cite{van2008visualizing}, as shown in Figure~\ref{fig:tsne} and Figure~\ref{fig:tsne2}. 
Although the inter-class distance and intra-class distance have changed after obfuscation, the distinguishability of the feature space topology is preserved.
Therefore, the model trained on obfuscated features can still achieve high accuracy close to the one trained on original features. 
But the class center of each identity has been shifted by obfuscation, therefore identity leakage has been reduced as long as the adversary does not know the obfuscation parameters.

\subsection{Across Diverse Datasets and Models of Varied Complexity.}
\label{subsec:eval_datasets}
% \subsection{(maybe) Multi datasets training}

% \begin{table*}[ht]
% \begin{center}
% \begin{tabular}{|l|l|l|l|l|l|l|}
% \hline
%                                                                             & XNN & HE & FL & DP & LDP & TEE \\ \hline
% \begin{tabular}[c]{@{}l@{}}Acceptable computation overhead\end{tabular} & Yes    &     & Yes  & Yes  & Yes   & Yes   \\ \hline
% \begin{tabular}[c]{@{}l@{}}Acceptable model accuracy\end{tabular}       & Yes    & Yes   & Yes  & Yes  &     & Yes   \\ \hline
% \begin{tabular}[c]{@{}l@{}}No access to original data\end{tabular}        & Yes    & Yes   & Yes  &    & Yes   & Yes   \\ \hline
% \begin{tabular}[c]{@{}l@{}}Low communication cost\end{tabular}          & Yes    & Yes   &    & Yes  & Yes   & Yes   \\ \hline
% \begin{tabular}[c]{@{}l@{}}Large model support\end{tabular}               & Yes    & Yes   & Yes  & Yes  & Yes   &     \\ \hline
% \begin{tabular}[c]{@{}l@{}}Cryptographic security\end{tabular}            &      & Yes   &    &    &     & Yes   \\ \hline
% \begin{tabular}[c]{@{}l@{}}Plaintext inference\end{tabular}            &      &   &  Yes  &  Yes  &   Yes  & Yes   \\ \hline
% \end{tabular}
% \end{center}
% \vspace{1mm}
% \caption{Comparison between private learning methods. XNN: our method, HE: Homomorphic Encryption, FL: Federated Learning, DP: Differential Privacy, LDP: Local Differential Privacy, TEE: Trusted Execution Environment}
% \label{tab:comparison_plm}
% \vspace{-3mm}
% \end{table*}

\noindent\textbf{Performance on Different Datasets.}
Table~\ref{tab:diff_dataset} presents the results of our experiments evaluating the performance of XNN on datasets with over three million images.
On larger datasets, such as VGGFace2 and MSRA, XNN demonstrates outstanding performance in balancing utility and privacy, surpassing the results presented in Table~\ref{tab:comparison_celeba} and~\ref{tab:comparison_webface} for CelebA and WebFace. 
Specifically, XNN achieves privacy protection that is nearly equivalent to random guessing while causing almost imperceptible degradation in utility.

These findings demonstrate the generalizability and effectiveness of XNN in protecting privacy for deep learning models, particularly on larger-scale datasets, which are more prevalent in practical applications. 
The ability of XNN to balance privacy and utility effectively makes it a promising solution for real-world scenarios where both aspects are critical.

\noindent\textbf{Evaluating Effect of Layers on Utility.}
The number of layers and their configuration can have a significant impact on the performance of the model, and often requires careful tuning and experimentation.
In order to verify that our proposed XNN can achieve good availability assurance in the face of different complexity networks, we evaluate the impact of network layers on availability on the CelebA dataset.

\begin{table}
	\centering
    \renewcommand\arraystretch{1.0}
    \resizebox{\columnwidth}{!}{
        \setlength{\tabcolsep}{10pt}
    	\begin{tabular}{c|cc|cc}
    		\toprule
            % \hline
    		 \multirow{2}{*}{Methods}     & \multicolumn{2}{c|}{VGGFace2} &  \multicolumn{2}{c}{MSRA} \\
    		% \cmidrule(r){2-3}
    		& Utils($\uparrow$) & ASR($\downarrow$) & Utils($\uparrow$) & ASR($\downarrow$) \\
    		\midrule
            % \hline
    		Vanilla+PE & 68.58 & 68.58  & 94.16 & 94.16     \\
            \rowcolor{gray!25} 
            XNN    & 67.56 & 0.15 & 93.56 & 0.17  \\
    		\bottomrule
            % \hline
    	\end{tabular}
    }
\vspace{1mm}
\caption{Performance (\%) on Different Datasets.}
\label{tab:diff_dataset}
\vspace{-1mm}
\end{table}

\begin{figure}
    \centering
  	\subfigure{
        \includegraphics[width=.45\textwidth]{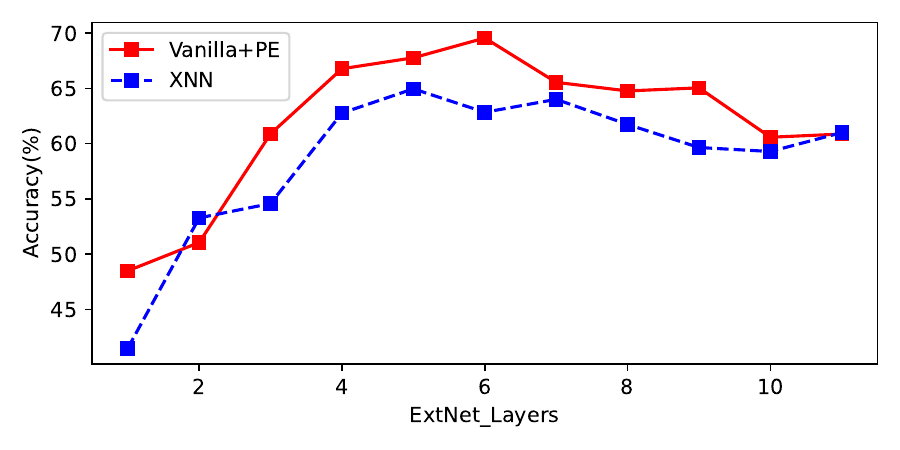}
		\label{fig:ext_layer}
	}
    \vspace{-3mm}
 	% \hspace{-3mm}
	\subfigure{
        \includegraphics[width=.45\textwidth]{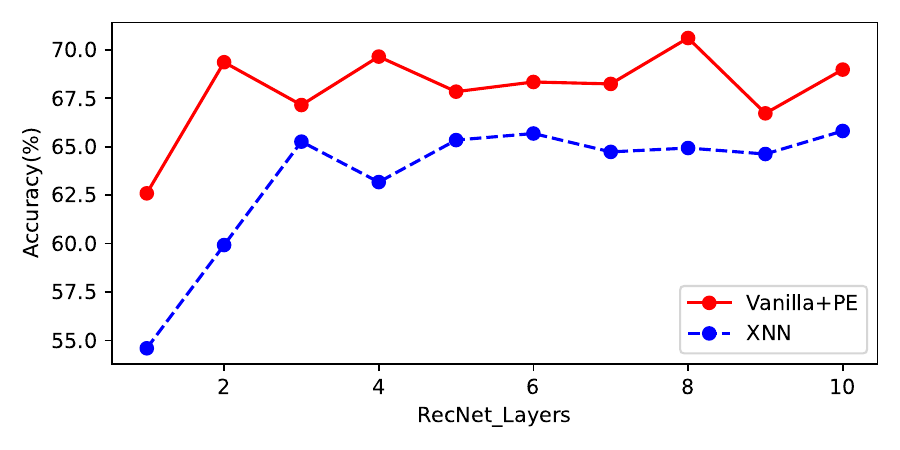}
			\label{fig:tail_layer}
	}
    \vspace{2mm}
    \caption{Effect of number of network layers on utility.}
    \label{fig:eval_layers}
    \vspace{-3mm}
\end{figure}

The results are shown in Figure~\ref{fig:eval_layers}, XNN achieved high accuracy close to the one of vanilla model across vsrying numbers of layers.
Specifically, in the case of RecNet, XNN showed an average reduction in utility of 4.55\% compared to vanilla models. In the case of ExtNet, this difference was smaller, reduced by 3.20\%. 
These results highlight the potential of XNN to maintain high utility even when the number of layers in the model is varied.
In addition, ExtNet achieved the highest performance when trained with 5 layers and RecNet exhibited optimal performance with 6 layers, which is the setting we used in the experiment above.

\section{Discussion}
\label{sec:Discussion}
\subsection{Comparison with other privacy approaches}

To reduce the identity leakage of a dataset, there are other classes of approaches, such as Homomorphic Encryption (HE)~\cite{acar2018survey}, Federated Learning (FL)~\cite{li2020federated}, Differential Privacy (DP)~\cite{dwork2008differential}, Local Differential Privacy (LDP)~\cite{cormode2018privacy} and Trusted Execution Environment (TEE)~\cite{sabt2015trusted}, each with strengths and limitations, and which provide different privacy guarantees.

The computation complexity of HE is quite large, which makes it impractical for heavy tasks such as neural network training.
FL requires that every dataset owner has sufficient computation power for model training, and almost all clients must be online every time the cloud platform wants to train a new model from the same dataset.
DP can prevent privacy leakage from the trained model, but the cloud platform still requires access to the original training data which is private.
LDP adds sufficient large noise to the dataset to protect privacy.
But the noise will severely decrease the accuracy of deep learning models.
TEE can only be applied to small deep-learning models due to its limited secure memory.

Compared with the above approaches, XNN has its unique advantage and limitation.
XNN obfuscates the dataset with random permutation and random matrix multiplication, which are not computationally expensive compared with HE.
Then the obfuscated feature is sent to the cloud platform for model training.
The cloud platform does not need access to the original training data like DP.
Also, XNN does not require that the dataset owner has the computation power for model training like FL.
Next, the cloud platform trains a recognition model on the obfuscated feature.
The size of the recognition model is not limited, like TEE.
XNN and LDP both modify the training dataset, but unlike LDP, XNN can still achieve accuracy close to the model trained on the original data.
However, XNN also has its limitation.
XNN is designed to reduce the identity leakage of the dataset.
It does not have cryptographic security.
And it can not make plaintext inferences, because it is only trained on the obfuscated features.
% As a summary, a comparison among the above methods is presented in Table~\ref{tab:comparison_plm}.

\subsection{Privacy risk under other adversarial settings}
Throughout this paper, we measure the privacy risk of XNN and XNN-d by the identity leakage under the expectation recognition attack described in Section~\ref{sec:EvaluationMethod}.
We assumed that the adversary is enhanced by the ability to access the gallery set of obfuscated features.
But in the real scenario, the dataset owner will not outsource the gallery set.
So we can expect a much lower identity leakage in the real setting.
However, using the same fixed obfuscation parameters for all samples in the dataset can be a potential weakness of XNN under adversarial attacks.
If several original-obfuscated pairs of features are leaked, the adversary may build an approximate inverse of the obfuscation and identify any other obfuscated features with the help of the inverse function.

\section{Conclusion}
\label{sec:Conclusion}

We proposed a method called XNN to reduce the identity leakage of the recognition dataset by obfuscation when the dataset owner outsources his data to the untrusted cloud platform for model training.
Experimental results show that XNN can effectively reduce identity leakage to almost 0 with an imperceptible loss of training accuracy.
Our work may inspire the development of novel obfuscation designs or applications for other machine-learning tasks with privacy concerns.

\bibliographystyle{named}
\bibliography{ijcai24}

\end{document}